# Power coupling

*D. Alesini*
LNF, INFN, Frascati, Italy

**Abstract**

Power coupling is the subject of a huge amount of literature and material since for each particular RF structure it is necessary to design a coupler that satisfies some requirements, and several approaches are in principle possible. The choice of one coupler with respect to another depends on the particular RF design expertise. Nevertheless some 'design criteria' can be adopted and the scope of this paper is to give an overview of the basic concepts in power coupler design and techniques. We illustrate both the cases of normal-conducting and superconducting structures as well as the cases of standing-wave and travelling-wave structures. Problems related to field distortion induced by couplers, pulsed heating, and multipacting are also addressed. Finally a couple of design techniques using electromagnetic codes are illustrated. The paper brings together pictures, data, and information from several works reported in the references and I would like to thank all the authors of the papers.

## 1      Introduction: basic concepts and coupler classification

Power couplers can generally be defined as networks designed to transfer power from an RF power source to a cavity, as illustrated in Fig. 1(a). Generally speaking, they are realized with slots, windows, or antennas that couple the electromagnetic (e.m.) field of the power transfer line to the cavity e.m. field. For this reason, a first classification (found in the literature) of couplers is between 'magnetic' or 'electric', depending on what type of field they couple. In some cases, nevertheless, this classification cannot be applied because they couple both the E and the H field, as discussed in the following.

Another possible classification of couplers can be between 'waveguide type' and 'coaxial type' couplers, depending on the particular geometry used to couple the power with the cavity field. Both types of coupler have advantages and drawbacks in terms of design, power handling capacity, and tunability. If necessary, special transitions between waveguide and coaxial lines (or vice-versa) can be integrated in the coupler itself (as an example, the power transfer line can be a waveguide and the final coupler can be coaxial); in this case a special transition has to be integrated in the coupler itself. The design of the particular coupler depends on the cavity type: normal-conducting (NC) or superconducting (SC), standing-wave (SW) or travelling-wave (TW) as shown in Fig. 1(b) where all principal combinations are given.

The design has to be oriented not only to the right power transfer but it has to include other important technical aspects. First of all, since, in general, transmission lines (coaxial or waveguide), are usually filled with gas, couplers have to incorporate vacuum barriers (RF windows) to preserve the cavity's operation under ultra-high vacuum conditions. This is, in particular, a critical point for SC cavities where vacuum contamination can seriously damage the structures. Moreover, in the case of SC cavities, an input coupler must serve as a low-heat-leak thermal transition between the room-temperature environment outside and the cryogenic temperature (from 2 to 4.5 K). Thermal intercepts and/or active cooling in the coupler design might be necessary. Finally, the coupler design cannot be independent of beam dynamics considerations. It introduces, in fact, an asymmetry in the electromagnetic field distribution which can deteriorate the beam quality. Special measures, such as double symmetric couplers or compensating stubs, may be required and integrated in the coupler itself.

In recent years the RF design of the couplers has been enormously aided by computer codes. These codes allow complete modelling of the field distribution in the couplers and cavities avoiding the use of the cut-and-try technique used in the past. The codes allow minimization in the design procedure of the pulsed heating and multi-pacting phenomena that can seriously limit coupler performance.

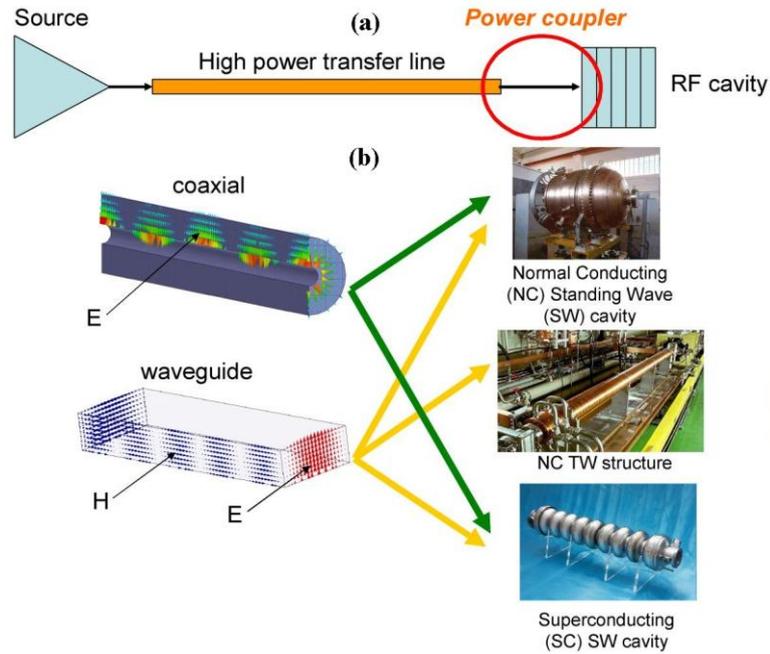

**Fig. 1:** (a) Conceptual sketch of a source–power transfer line–coupler–cavity system; (b) schematic representation of possible combinations of coaxial and waveguide couplers with different types of cavity

## 2   Coupling to standing-wave cavities

Standing-wave cavities can be excited using slots on waveguides or loops and antennas in coaxial couplers. The general problem of coupling between a power transfer line and a cavity is not trivial and only perturbative analytical approaches can be followed [T1–T4]. The problem is that coupling slots or antennas change the cavity and waveguide geometry and, when the field into the cavity starts to increase, the cavity itself radiates into the waveguide through the same slot or antenna. In steady state, and at a given frequency of operation, there is equilibrium between the power flowing into the cavity, that dissipated into the walls, and that radiated from the cavity back into the power transfer line. The analytical, exact solution of such a problem is in general impossible.

Before introducing the approximated analytical formulae that allow the modelling of this coupling, let us start with an intuitive and qualitative approach to the subject.

### 2.1   Slots on waveguides

Let us consider, for the sake of simplicity, the case of the excitation of the $TM_{010}$ accelerating mode of a simple pillbox cavity by a waveguide coupled to the cavity with a slot. Looking at the magnetic field lines of a short-circuited waveguide and cavity [Fig. 2(a)], it is straightforward to note that there are two possible simple solutions to couple the magnetic field of the waveguide with the magnetic field lines of the mode and they are given in Figs. 2(b) and 2(c).

If the linear dimensions of the aperture are small compared with the wavelength, it is possible to use a perturbative approach in which the aperture is equivalent to a magnetic dipole ($\vec{M}_{slot}$), whose dipole moment is proportional to the tangential magnetic field on the slot ($\vec{H}_{slot}$). This dipole is the source of the field into the cavity and the source of the field radiated back from the cavity into the waveguide. As discussed in the next paragraphs the final amplitude of the mode into the cavity is proportional to the scalar product $\vec{H}_{slot} \cdot \vec{H}_{cavity}$, where $\vec{H}_{cavity}$ is the magnetic field into the closed cavity (without the slot). From the formula it is easy to note that, in order to have mode excitation, both $\vec{H}_{cavity}$ and $\vec{H}_{slot}$ must be different from zero and non orthogonal.

It is also straightforward to note that the excitation of the field inside the cavity, in the case of Fig. 2(c), can be varied (for a fixed coupling slot aperture) by changing the distance $d$ between the short-circuit plane and the slot itself. The distance $d$, in fact, varies the amplitude of the final $\vec{H}_{slot}$.

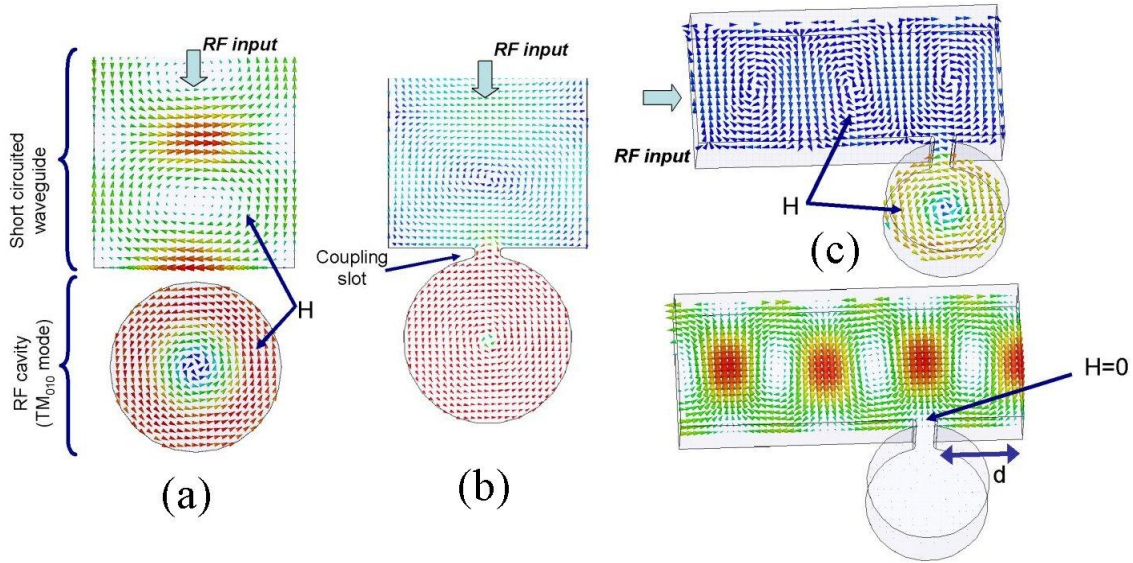

**Fig. 2:** Magnetic coupling between a waveguide and a SW cavity: (a) magnetic field lines on a short-circuited waveguide and cavity operating on the $TM_{010}$ mode; (b) first solution of magnetic coupling between the waveguide and the cavity; (c) second solution for magnetic coupling (the distance $d$, in this case, allows changing the excitation of the mode)

## 2.2 Loops on coaxial

Magnetic coupling can be also realized using loops, as shown in Fig. 3. In this case also it is possible, for small loop dimensions, to use a perturbative approach. The loop excites into the cavity a magnetic dipole whose intensity is proportional to the loop area and coaxial input power, while the amplitude of the excited mode is proportional to the scalar product between this magnetic dipole and the magnetic field of the excited mode in the loop region ($\vec{M}_{loop} \cdot \vec{H}_{cavity}$).

It is straightforward to note that, also in this case, the amplitude of the mode can be varied (for a fixed loop dimension) by changing the loop orientation.

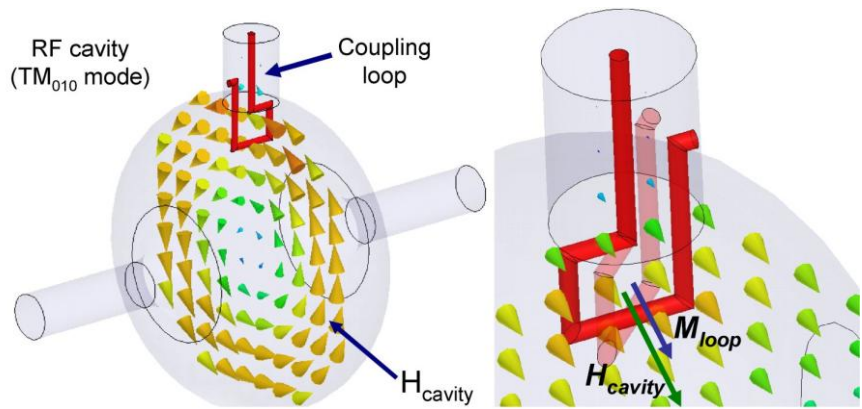

**Fig. 3:** Magnetic coupling between a loop and a SW cavity: the excitation of the mode can be varied by changing the orientation of the loop

## 2.3 Antenna on coaxial

Electric coupling can be realized by an antenna in the case of coaxial couplers. The typical situation is illustrated in Fig. 4. In this case the inner conductor of the coaxial is inserted into the cavity and the electric current on its surface is coupled with the electric field of the mode in the cavity. If the linear dimensions of the antenna are small with respect to the wavelength, it is possible to use, also in this case, a perturbative approach in which the inner is equivalent to an electric dipole ($\vec{P}_{antenna}$), whose dipole moment is proportional to the tangential density current on the inner itself. This dipole is the source of the field into the cavity and, vice-versa, when the cavity field starts to increase, it is the source of the field radiated back into the coaxial from the cavity. As illustrated in the following paragraph, in this case the amplitude of the excited mode is proportional to the scalar product $\vec{P}_{antenna} \cdot \vec{E}_{cavity}$.

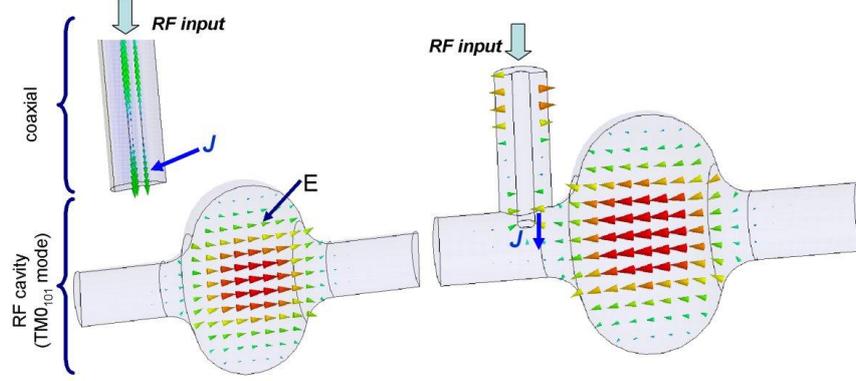

**Fig. 4:** Electric coupling between a coaxial and a SW cavity: the excitation of the mode can be varied by changing the penetration of the inner

## 2.4 Basics of the theory of coupling to standing-wave cavities

The theory of cavity excitation has been extensively treated using several equivalent approaches. Without entering into the details of the rigorous treatments that can be found in the previous references, we summarize in the following the basic results.

Let us consider first the case of a single, closed, metallic cavity without external coupling. The resonant modes in such a cavity can be expanded in a series. To have a complete expansion we need both solenoidal modes (with zero divergence and non-zero curl) and irrotational modes (with zero curl but non-zero divergence). In most practical cases we deal with the first type of modes and, in the frequency domain, we can consider the following expansion of modes in a closed, ideal (without losses), metallic cavity:

$$\vec{E}e^{j\omega t} = \sum_n e_n \vec{E}_n e^{j\omega t} \quad ,$$
$$\vec{H}e^{j\omega t} = \sum_n h_n \vec{H}_n e^{j\omega t} \quad , \tag{1}$$

where $e_n$ and $h_n$ are the mode amplitudes (complex in general and frequency-dependent) while the eigenfunctions $\vec{E}_n, \vec{H}_n$ are solutions of the following equations:

$$\begin{cases} \nabla^2 \vec{E}_n + k_n^2 \vec{E}_n = 0 \\ \vec{n} \times \vec{E}_n = 0 \\ \nabla \cdot \vec{E}_n = 0 \end{cases} \quad \begin{cases} \nabla^2 \vec{H}_n + k_n^2 \vec{H}_n = 0 & \text{in the volume of the cavity } V \quad , \\ \vec{n} \cdot \vec{H}_n = 0 & \text{on the surface of the cavity } S \quad , \\ \nabla \cdot \vec{H}_n = 0 & \text{in } V \text{ and on } S \quad . \end{cases} \tag{2}$$

The first equations are the well-known Helmoltz equations and $k_n$ are the eigenvalues of the problems, related to the resonant frequency of the mode by the usual relation $\omega_n = k_n(\mu\varepsilon)^{-1/2}$. The eigenfunctions and the eigenvalues can be chosen real and orthonormal. This means that

$$\int_V \vec{E}_n \cdot \vec{E}_m = \delta_{nm} \qquad \int_V \vec{H}_n \cdot \vec{H}_m = \delta_{nm} \quad , \tag{3}$$

where the Kronecker delta $\delta_{nm} = 0$ if $n \neq m$ and $\delta_{nm} = 1$ if $n = m$. The relationship between the eigenfunctions $\vec{E}_n$ and $\vec{H}_n$ and between $e_n$ and $h_n$ are given by Maxwell's equations. It can be demonstrated that

$$\nabla \times \vec{E}_n = k_n \vec{H}_n \qquad \nabla \times \vec{H}_n = k_n \vec{E}_n \quad . \tag{4}$$

It is easy to demonstrate that in a source-free cavity with perfectly conducting walls, $e_n \neq 0$ if and only if $\omega = \omega_n$ and that

$$h_n = j\sqrt{\frac{\varepsilon}{\mu}} e_n \quad . \tag{5}$$

In this case, therefore, the field can be expanded in the following sum:

$$\vec{E} e^{j\omega t} = \sum_n e_n \vec{E_n} e^{j\omega_n t} \quad ,$$
$$\vec{H} e^{j\omega t} = \sum_n h_n \vec{H_n} e^{j\omega_n t} \quad . \tag{6}$$

The case of wall losses can be treated using a perturbative approach starting from the case of a source-free cavity with lossy walls. In this case it is possible to demonstrate that the coefficients $e_n$ and $h_n$ are different from zero if and only if $\omega$ satisfies this equation:

$$\omega = \omega_n \left(1 - \frac{1}{2Q_{0n}}\right) + j\underbrace{\frac{\omega_n}{2Q_{0n}}}_{\alpha} \quad , \tag{7}$$

where the quality factor $Q_{0n}$ is defined as:

$$Q_{0n} = \omega_n \frac{W}{P_d} \qquad P_d = \frac{1}{2} R_s \int\limits_{\substack{cavity \\ walls}} \left|\vec{H_n}\right|^2 dS \qquad W = \frac{1}{2}\mu \int\limits_{\substack{cavity \\ volume}} \left|\vec{H_n}\right|^2 dV = \frac{1}{2}\varepsilon \int\limits_{\substack{cavity \\ volume}} \left|\vec{E_n}\right|^2 dV \tag{8}$$

where $P_d$ and $W$ are the average dissipated power into the cavity walls and the average electromagnetic energy stored in the cavity. In Eq. (8) $R_s$ is the surface resistivity, defined as:

$$R_s = \sqrt{\frac{\omega_n \mu}{2\sigma}} \qquad \sigma = \text{cavity wall conductivity} \quad . \tag{9}$$

Equation (7) shows that the frequency of free oscillation slightly differs from the no-loss resonant frequency and, in addition, there is a damping constant $\alpha$.

Let us now consider the case of a cavity coupled externally with a waveguide or a coaxial. These cases are illustrated in the previous Figs. 2–4. Following, also in this case, a perturbative approach, the excitations of a mode in a cavity can be modelled by an equivalent electric ($\vec{J}$) or magnetic ($\vec{J_m}$) density current representing the sources of the modes. The equivalent magnetic sources are, for example, the magnetic field on a coupling slot between the waveguide and the cavity (Fig. 2) and the magnetic field generated by a loop coupled with a cavity (Fig. 3), while the equivalent electric sources are the currents on a small antenna coupled with the cavity (Fig. 4).

It can be demonstrated that, if the frequency of the external excitation ($\omega$), is very near to a particular mode resonant frequency $\omega_n$, all coefficients in the expansion (1) are small with respect to the coefficients $e_n$, $h_n$ related to the $n$-mode and these latter coefficients are given by

$$h_n \cong \frac{k_n \int\limits_{\substack{electric \\ coupling \\ region}} \vec{J} \cdot \vec{E_n} - j\omega\varepsilon \int\limits_{\substack{magnetic \\ coupling \\ region}} \vec{J_m} \cdot \vec{H_n}}{k_n^2 - k^2\left(1 + \frac{1-j}{Q_{0n}}\right)} \cong -\frac{j\frac{Q_{0n}}{k_n} \int\limits_{\substack{electric \\ coupling \\ region}} \vec{J} \cdot \vec{E_n} + \frac{Q_{0n}}{\omega_n \mu} \int\limits_{\substack{magnetic \\ coupling \\ region}} \vec{J_m} \cdot \vec{H_n}}{1 + jQ_{0n}\underbrace{\left(\frac{\omega}{\omega_n} - \frac{\omega_n}{\omega}\right)}_{\delta}}$$

$$e_n \cong -\frac{j\omega\mu \int\limits_{\substack{electric \\ coupling \\ region}} \vec{J} \cdot \vec{E_n} + k_n \int\limits_{\substack{magnetic \\ coupling \\ region}} \vec{J_m} \cdot \vec{H_n}}{k_n^2 - k^2\left(1 + \frac{1-j}{Q_{0n}}\right)} \cong \frac{-\frac{Q_{0n}}{\omega_n \varepsilon} \int\limits_{\substack{electric \\ coupling \\ region}} \vec{J} \cdot \vec{E_n} + j\frac{Q_{0n}}{k_n} \int\limits_{\substack{magnetic \\ coupling \\ region}} \vec{J_m} \cdot \vec{H_n}}{1 + jQ_{0n}\delta} \quad . \tag{10}$$

The second equalities are valid since, in general $Q_{0n} \gg 1$ and $\omega$ in the numerator can be approximated with $\omega_n$. In other words, the amplitudes of the electric and magnetic field excited into the cavity have a typical resonant behaviour as a function of the frequency of the excitation and are different from zero only if $\omega \approx \omega_n$.

If in the coupling region the electric and magnetic fields of the modes are sufficiently small and can be approximated by their values in the centre of the coupler region and the integrals of the density currents can be substituted by their equivalent dipole moments, Eq. (10) becomes

$$h_n \cong \frac{j\omega k_n \vec{P} \cdot \vec{E}_n + k^2 \vec{M} \cdot \vec{H}_n}{k_n^2 - k^2\left(1 + \frac{1-j}{Q_{0n}}\right)}$$

$$e_n \cong -\frac{-\omega^2 \mu \vec{P} \cdot \vec{E}_n + j\omega\mu k_n \vec{M} \cdot \vec{H}_n}{k_n^2 - k^2\left(1 + \frac{1-j}{Q_{0n}}\right)} \quad . \tag{11}$$

For a given coupler geometry and SW cavity, it is, in general, very difficult to find an analytical expression for Eqs. (10)–(11) and only approximated treatments can be done. The previous considerations allow one to obtain, nevertheless, some important practical results as illustrated in the following section.

## 2.5 Equivalent circuit of the coupling between a standing-wave cavity and a power transfer line and definition of the coupling coefficient

Equations (10) and (11) state that, at a given frequency, the mode excitation is proportional to the equivalent source density currents ($\vec{J}, \vec{J}_m$) or dipole moments ($\vec{P}, \vec{M}$) and these sources are proportional, obviously, to the field in the coupler region. As usual, we can associate to the field into the cavity and to the field into the waveguide (or coaxial) an equivalent voltage and current proportional to the corresponding electric and magnetic fields or, more precisely, to their integrals over a given path.

Let us consider, for example, the case represented in Fig. 2(b) and let us indicate with $V_{cav}$ the equivalent cavity voltage and with $V_{waveg}$ and $I_{waveg}$ the equivalent voltage and current into the waveguide in the coupler region, given by the superposition of an incident wave ($V^+$) and a reflected one ($V^-$). From Eqs. (10)–(11) the cavity voltage is proportional to the magnetic field in the hole aperture and therefore to $I_{waveg}$ and, in the limit of a small coupling hole, the reflected wave $V^-$ will be given by the superposition of the reflected wave from the short-circuited plane at the end of the waveguide and the radiated field from the cavity through the coupling hole. Let us write these relationships:

$$\begin{cases} V^- = -V^+ + \dfrac{V_{cav}}{n} \\ V_{cav} = I_{waveg}\dfrac{R/n}{1+jQ_{0n}} = \left(\dfrac{V^+ - V^-}{Z_0}\right)\dfrac{R/n}{1+jQ_{0n}\delta} \end{cases}, \tag{12}$$

where we have taken into account Eqs. (10)–(11) in the cavity voltage expression and where we have introduced the waveguide characteristic impedance $Z_0$ and two quantities $n$ and $R$. The first one is an adimensional quantity that relates the cavity voltage to the radiated voltage from the cavity into the waveguide; in other words if, at a certain time, we switch off the excitation, the cavity radiates into the waveguide and the ratio between the radiated voltage and the cavity voltage is $1/n$. The other quantity is $R/n$ (ohm) and it relates the waveguide magnetic current to the maximum amplitude of the cavity voltage, when the cavity is excited perfectly on resonance ($\delta = 0$).

From Eqs. (12) it is easy to understand that the coupling between a waveguide and a cavity can be modelled by the circuit given in Fig. 5. The cavity is modelled with an equivalent lumped RLC circuit, the RF source by the current generator, while the transformer models the coupler.

In this equivalent circuit the RLC circuit must have the same quality factor as the resonant mode and the same resonant frequency. This means that

$$\omega_n = \frac{1}{\sqrt{LC}}; \quad Q_o = R\sqrt{\frac{C}{L}} \qquad (13)$$

In Eq. (13) and from now on we consider just a single resonant mode and we indicate with $f_{res}$ ($\omega_{res}$) and $Q_0$ its resonant frequency and quality factor.

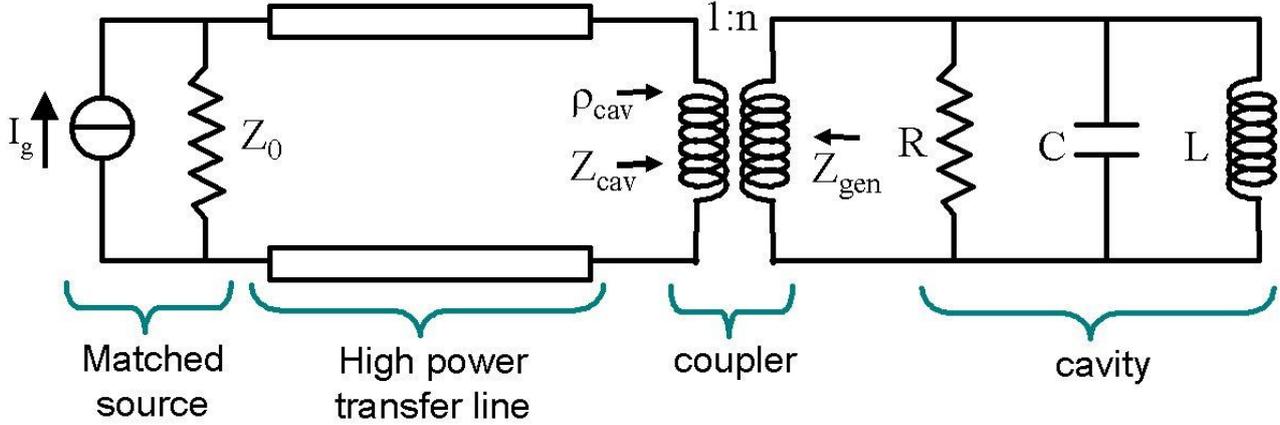

**Fig. 5:** Equivalent circuit of a cavity coupled with an RF source

From the generator point of view the cavity can be seen as a terminal impedance with the following value:

$$Z_{cav} = \frac{R/n^2}{1 + jQ_0\delta} \qquad (14)$$

On the other hand, the energy stored in the cavity ($W$) is dissipated into the cavity walls ($P_{cav}$) and is also radiated through the coupler and dissipated into the matched load of the generator ($P_{ext}$). Therefore the generator can be seen from the cavity point of view as an extra-load and we can define the following quantities:

a) The loaded quality factor that takes into account the total dissipated power:

$$Q_L = \frac{\omega_{res} W}{\underbrace{P_{cav} + P_{ext}}_{P_T}} \qquad (15)$$

While $Q_0$ is called the unloaded quality factor, and $P_T$ is the total average power dissipated into the cavity and into the external load.

b) The external quality factor $Q_E$ defined as

$$Q_E = \frac{\omega_{res} W}{P_{ext}} \qquad (16)$$

The quality factors are related by the following relationship:

$$\frac{1}{Q_L} = \frac{1}{Q_0} + \frac{1}{Q_E}. \qquad (17)$$

c) The coupling coefficient $\beta$ is defined as

$$\beta = \frac{P_{ext}}{P_{cav}} = \frac{Q_0}{Q_E} = \frac{R}{n^2 Z_0} \qquad (18)$$

From the previous relations it is possible to obtain that

$$Q_L = \frac{Q_0}{1+\beta} \qquad (19)$$

The reflection coefficient seen by the source under the assumption of zero length power line is then

$$\rho_{cav} = \frac{\beta - 1 - jQ_0\delta}{\beta + 1 + jQ_0\delta}; \qquad \rho_{cav}\big|_{f=f_{res}} = \frac{\beta - 1}{\beta + 1} \quad . \tag{20}$$

From the previous equations it is straightforward to note that the coupling β fixes the reflection coefficient at the input port, the resonance bandwidth, and the ratio between the power dissipated into the cavity and external load. It plays an important role in the design of a cavity, and the choice of its value depends on several parameters such as the beam loading and the available input power from the source. It is possible, in general, to change it by changing the geometry of the coupler as given in Fig. 6. The particular case of β = 1 is called critical coupling and, in this case, in the steady-state regime, we have no reflected power to the generator because there is a perfect destructive interference between the power reflected from the waveguide (or coaxial) termination and that radiated back by the cavity through the hole. The case with β < 1 is called undercoupling while the case with β > 1 is called overcoupling. As an example the reflection coefficient at the coupler input port, near the cavity resonant frequency, is given in Fig. 7 with the assumption $f_{res}$ = 1 GHz, $Q_0$ = 40 000 and for three different values of the coupling.

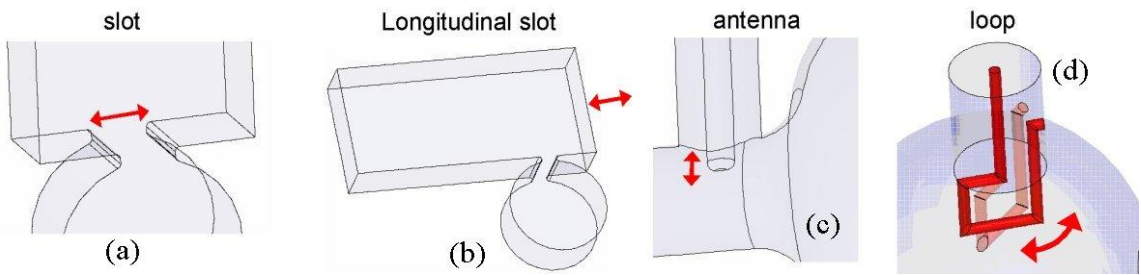

**Fig. 6:** Typical geometry modifications to change the coupling: (a) slot aperture; (b) short-circuit terminal plane; (c) antenna penetration; (d) loop orientation

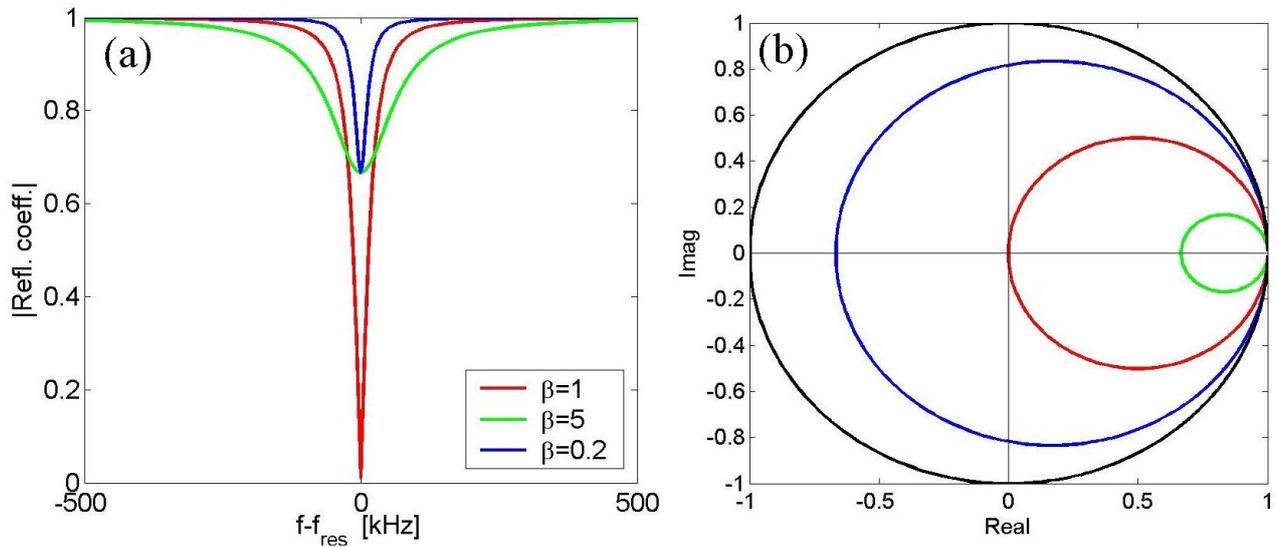

**Fig. 7:** Reflection coefficient at the coupler input port for a cavity resonating at $f_{res}$ = 1 GHz with $Q_0$ = 40 000 for three different value of the coupling coefficient: (a) absolute value; (b) complex plane

In the design procedure (by, for example, an electromagnetic code) it is possible to calculate and tune the coupling coefficient following these two steps:

a)  Establish if we are under-coupled or over-coupled. To do this it is sufficient to look at the reflection coefficient at the coupler input port (as a function of frequency) in the complex plane. Out of resonance the reflection coefficient has an absolute value equal to 1 and, therefore, it describes a circle with radius equal to 1. At resonance it describes a circle towards the origin of the complex plane. It is easy to demonstrate that, if the circle includes the origin of the complex plane we are overcoupled, if not we are undercoupled. If the circle crosses the origin we are in critical coupling.

b)      Once we have established if we are over- or undercoupled, we can calculate the coupling by the formulae simply derived from Eqs. (20):

$$\beta = \frac{1+|\rho_{cav}|}{1-|\rho_{cav}|} \quad \text{(overcoupled)},$$

$$\beta = \frac{1-|\rho_{cav}|}{1+|\rho_{cav}|} \quad \text{(undercoupled)}.$$

(21)

Here $|\rho_{cav}|$ is the absolute value of the reflection coefficient at resonance.

## 2.6  Advantages and disadvantages of waveguide and coaxial couplers

Table I illustrates the main advantages and disadvantages of waveguide and coaxial couplers (well discussed in [S4]).

**Table 1:** Advantages and disadvantages of waveguide and coaxial couplers

| Type | Advantages | Drawbacks |
|---|---|---|
| Waveguide | Higher power handling capability | Larger size |
| | Easier to cool | More difficult to make variable |
| | Higher pumping speed | Bigger heat leak |
| | Simpler design: it does not require a transition between the waveguide, which usually carries the output power of the RF sources, and the cavity interface. | |
| | Lower attenuation | |
| Coaxial | Easier to make variable | More complicated design |
| | More compact | Worse power handling |
| | Smaller heat leak | Higher attenuation |
| | Multipacting levels easier to manage | Smaller pumping speed |
| | | More difficult to cool (coaxial inner) |

Couplers on waveguide are preferred at high frequency (>1 GHz) and for high-gradient and high-power structures likes TW linacs and normal-conducting RF guns. Couplers on coaxial are preferred for low-frequency structures (<1 GHz) and when the coupling has to be varied, for example, for superconducting structures (due also to the smaller heat leak). The analysis of the coupler in the following sections will clarify some of these features.

## 3  Input couplers for normal-conducting standing-wave cavities: an analysis of the critical features

We will illustrate the main critical points in the design of input couplers for normal-conducting standing-wave cavities looking at some practical examples.

### 3.1  Normal-conducting standing-wave cavities: RF guns

Normal-conducting RF guns are the first stage of acceleration of e$^-$ in the linacs for FELs [G1]. They are, in general, 2–3 cell SW accelerating structures operating on the π mode at frequencies of the order of a few GHz. Required coupling coefficients β are between 1 and 2. The operation of RF guns is, in general, pulsed with high peak input power (from a few up to 15 MW), pulse length from a few μs up to one ms, and repetition rate from a few Hz up to kHz. Because of the high-frequency, high-accelerating gradient (~50–100 MV/m), high-input-power operation, and fixed coupling, these structures are fed by slots on waveguides, as illustrated in Fig. 8. The main critical points in the design, realization, and operation of such couplers are related to the field distortion introduced by couplers and pulsed heating in the coupler region.

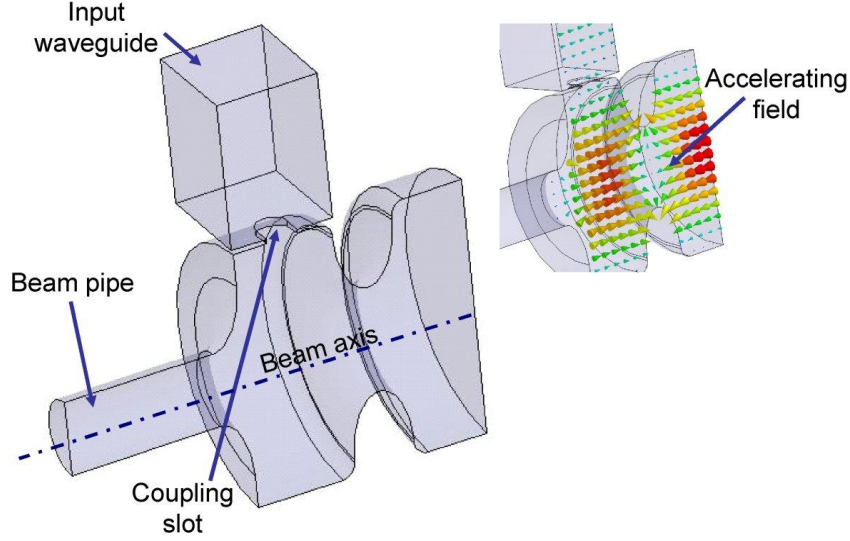

**Fig. 8:** Sketch of an RF gun with input coupler

### 3.1.1 *Field distortion*

Because of the relatively low energy of the electron beams (from 0 up to few MeV) an excellent uniformity is required for the accelerating field to preserve the beam quality. 'Standard' coupling slots introduce a distortion in the field distribution and multi-pole components of the field can appear and affect the beam dynamics [BD1]. The main mechanism that guides such field distortion is sketched in Fig. 9 where the magnetic field lines of a pure pillbox cavity and those of a cavity with a single and a double input coupler are shown. It is quite clear that in the first case the coupler introduces both a distortion of the field in terms of dipole component on axis and higher order components (quadrupole, sextupole, etc.), while in the case of a symmetric feeding, the odd magnetic field components of the field (like dipole, sextupole, etc.) are suppressed but we still can have even components (like quadrupoles, octupoles, etc.). More precisely we can develop the magnetic field near the beam axis as follows:

$$B_\phi(r,\phi,z) \cong A_o(z)r + \sum_{n=1}^{\infty} A_n(z)\cos(n\phi)r^{n-1} \quad , \tag{22}$$

where $z$ is the longitudinal coordinate, $r$ is the radial one and $\varphi$ is the azimuthal angle, as illustrated in Fig. 9. The $A_n$ components are, in general, complex functions and depend on the longitudinal coordinate $z$. The dipole component corresponds to the term $A_1$, the quadrupole one to the term $A_2$, and so on. In the case of a symmetric feeding only the even components remain.

The main techniques to minimize such field distortion are shown in Fig. 10. The simple way to partially compensate the dipole field is to introduce a symmetric compensating slot opposite to the RF input one [Fig. 10 (a)]. The opposite slot is not fed by RF power and can be used, for example, by pumping the structure (as reported in the same figure where the picture of the SLAC-UCLA-BNL gun is shown [G2]). In this case the field is partially compensated in the sense that the even components still remain and also the odd components have a residual amplitude on axis, due to the fact that the power is still entering from one side only.

To suppress the odd components the straightforward way is to use a symmetric feed [Fig. 10 (b)]. In this case a double feed is needed and it can be realized with a splitter that splits the main input power into two branches. Of course this system introduces a further complication and an increase in the cost of the complete realization due to the presence of the splitter itself. Nevertheless, also in this case the quadrupole component still survives and the only way to suppress it is by the use of a deformed profile of the cells (racetrack profile [G4, G5]) [as given in Fig. 10 (c)]. This cell deformation complicates further the machining of the cell that cannot be fabricated by lathe but needs a computer-controlled milling machine.

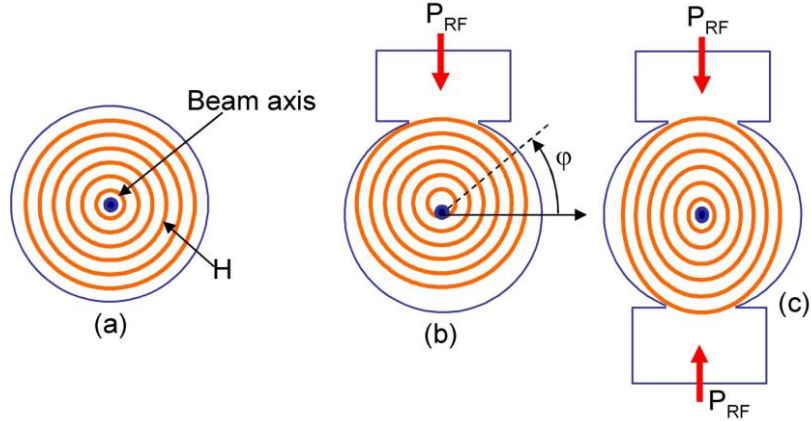

**Fig. 9:** Sketch of the magnetic field lines in a pure pillbox cavity (a), in a cavity with single input coupler (b), and in a cavity with double input coupler (c)

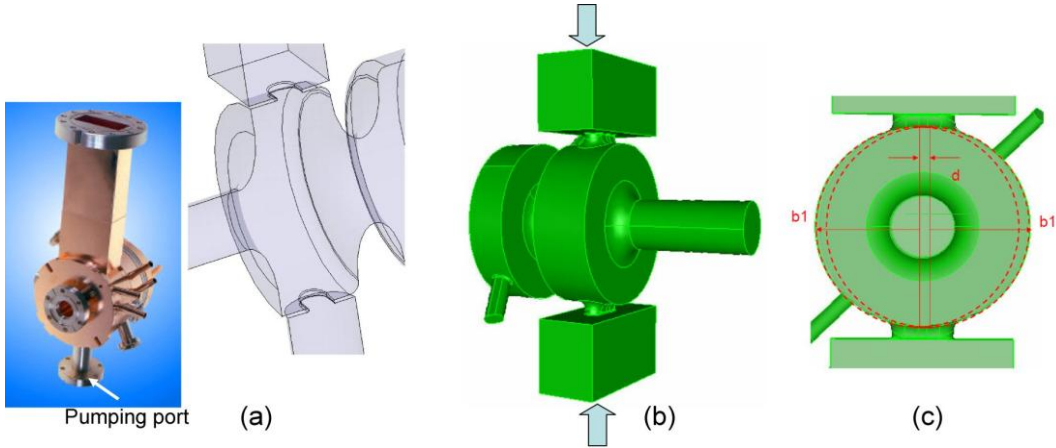

**Fig. 10:** Main techniques to minimize the field distortion introduced by the coupler: (a) compensating slot ([G2]); (b) dual feed; (c) dual feed and deformed internal profile of the cell

### 3.1.2 *High magnetic field in couplers and pulsed heating*

As mentioned before, RF guns are structures fed by typical RF pulses of several MW. High peak electric field (>100 MV/m) can be easily reached and breakdown phenomena can occur. These can be driven also by a high magnetic field in the coupler region, as reported in [B0, B1]. In fact, RF power enters the structure through the slots, the surface currents flow along the edges of the slots, and these edges are a place where local currents are significantly amplified. The high currents can give localized losses that can create hot spots and drive breakdown phenomena damaging the coupler itself.

The pulse temperature rise $\Delta T$ can be calculated using a 1D model [B2] and is given by

$$\Delta T = \frac{\left|H_\parallel\right|^2 \sqrt{t}}{\sigma \delta \sqrt{\pi \rho' c_\varepsilon k}} \quad , \tag{23}$$

where $t$ is the pulse length, $\sigma$ is the electrical conductivity, $\delta$ the skin depth, $\rho'$ the density, $c_\varepsilon$ the specific heat, and $k$ the thermal conductivity of the metal. For copper it is possible to use the following formula:

$$\Delta T [°C] = 127 \left|H_\parallel [\text{MA/m}]\right|^2 \sqrt{f_{RF}[\text{GHz}]} \sqrt{t[\mu s]} \quad . \tag{24}$$

As a general experimental rule, if this pulsed heating exceeds ~100 °C serious damage to the coupler region has a high probability of occurrence, below 50 °C damage to the couplers is practically avoided while in the interval 50–100 °C there is some probability of coupler damage.

The possible solutions to reduce the high magnetic field in the couplers are shown in Fig. 11. The strategy is to increase the curvature radius of the coupler slot. In particular the so-called *z*-coupling was the final solution adopted at SLAC for the LCLS gun [G4, G5]. Figure 11 (b) gives, for example, the pulsed heating as a function of the internal curvature radius of the *z*-coupler itself with the parameters of the LCLS gun given in the figure caption.

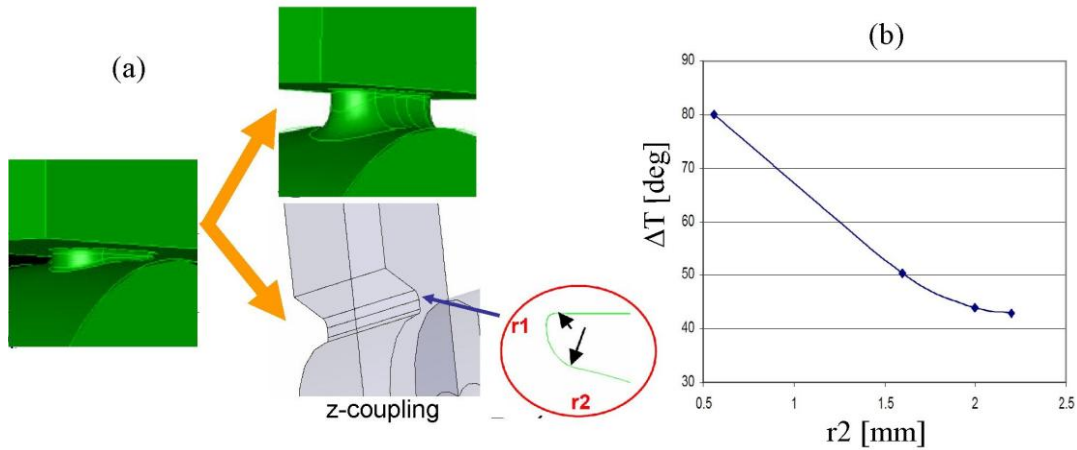

**Fig. 11:** (a) Possible solutions to reduce the high magnetic field in the couplers; (b) pulsed heating as a function of the internal radius of curvature of the *z*-coupler itself with the parameters of the LCLS gun [G5] ($f_{RF}$ = 2.856 GHz, $t$ = 3 μs, $E_{cathode}$ = 120 MV/m)

### 3.1.3 *On-axis coaxial coupling*

To simultaneously suppress the multipole components of the field and to strongly reduce the magnetic field in the coupler region, a new type of coupler for RF guns has been proposed and implemented [GA0, GA1, GA2]. The coupling is coaxial and on-axis coaxial as shown in Fig. 12. The $TE_{11}$ mode of the waveguide is converted into a TEM mode of a coaxial line (in the so-called door-knob transition) and the coaxial line excites, on-axis, the accelerating field in the gun. The excitation is both electric and magnetic since on-axis there are both field components. All mode asymmetries in the door-knob transition are suppressed in the coaxial line and the excitation of the cavity is purely 2D symmetric without multipole components [BD2].

It is also easy to verify that, in this case, the magnetic field in the coupler region (and therefore the pulsed heating) is strongly reduced since there are no sharp edges or coupling slots.

On the other hand, this type of coupler complicates the coupler design and gun realization. The guns of the Free Electron Laser in Hamburg (FLASH) and of the Photoinjector Test Facility at DESY in Zeuthen (PITZ) are based on this principle.

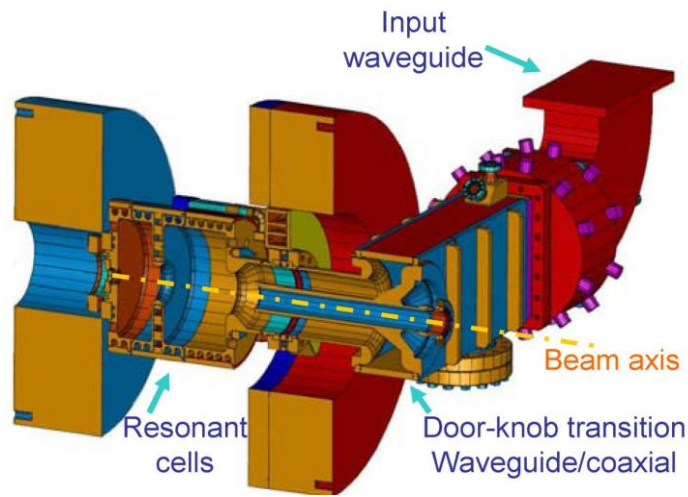

**Fig. 12:** On-axis coaxial coupling for RF guns (FLASH RF gun [GA1])

## 3.2 Coaxial-loop couplers for standing-wave, normal-conducting cavities: DAΦNE cavity

An example of the coaxial-loop coupler is the case of the DAΦNE cavity at LNF-INFN, Frascati, Italy [G7]. A sketch of the cavity and a detail of the loop mechanical drawing are given in Fig. 13. The cavity operates at 368 MHz with a maximum input power of 100 kW. The coupler integrates a transition between the rectangular waveguide and the coaxial. The loop can be rotated, changing the coupling coefficient from 0 to 10, and water cooling is integrated in the loop itself.

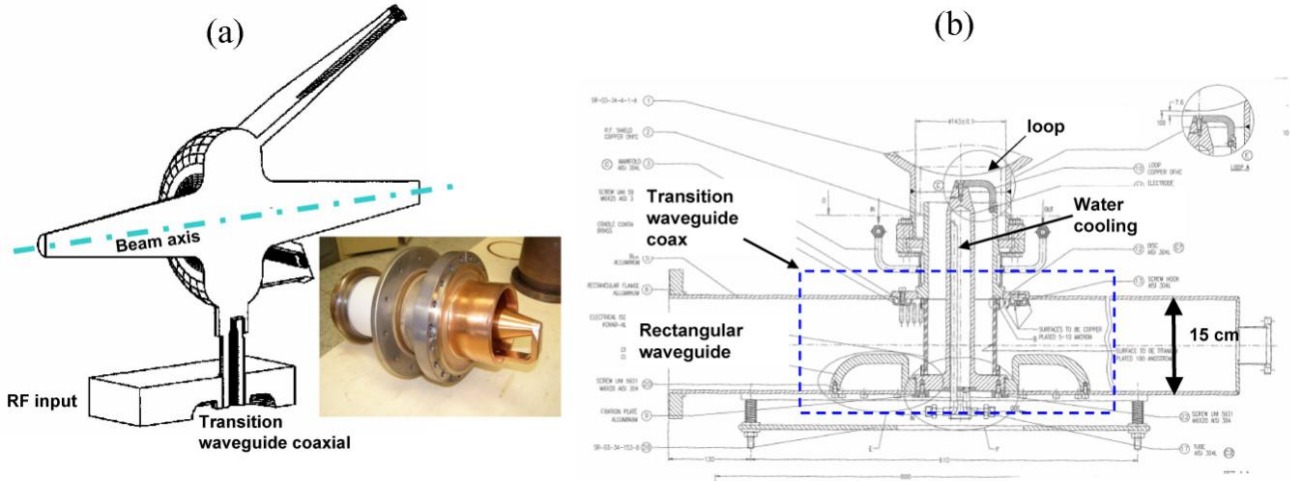

**Fig. 13:** (a) Sketch of the DAΦNE cavity and picture of the coaxial-loop coupler; (b) detail of the coupler mechanical drawing

## 4 Input couplers for superconducting standing-wave cavities: an analysis of the critical features

Superconducting cavities are used for both electron and heavy particle acceleration in an extremely wide range of applications. Owing to the low surface resistance (about 10 nΩ at 2 K), their quality factors may exceed $10^{10}$; this means that only a tiny fraction of the incident RF power is dissipated in the cavity walls and most of it is either transferred to the beam or reflected into the source load. The input power can be from a few kW up to a few MW. Superconducting cavities allow CW operation and in pulsed regime and this affects the coupler design. The transient nature of pulsed power introduces thermal and mechanical stress in critical components that have to be evaluated. On the other hand, CW operation requires special solutions for power handling and coupler cooling. Several excellent reviews on couplers for superconducting cavities can be found in the literature [S1–S4].

In the case of superconducting cavities the main function of the coupler, providing power, has to be merged and integrated with other important requirements like vacuum, cryogenic and coupling tunability.

Both waveguides and coaxial couplers can be used for superconducting cavities. In both cases the coupling is realized on one side of the cavity structure and the profile of the cells is a purely symmetric 2D. This is done to avoid the opening of slots on the accelerating cells that can create magnetic hot spots in the cells themselves with additional design complications and potential problems in the cavity operation at high gradient. In coaxial-type couplers the coupling strength depends on the longitudinal location and size of the coupling port and can be varied by changing the insertion of the inner.

As an example, the mechanical drawing of the coupler for the TESLA Test Facility (TTF) cavities [S7, S8] is given in Fig.14. The TESLA cavity is a nine-cell standing-wave structure of about 1 m length whose lowest TM mode resonates at 1300 MHz. The cavity is made of solid niobium and is cooled by superfluid helium at 2 K. Each nine-cell cavity is equipped with its own titanium helium tank and with a coaxial RF power coupler capable of transmitting more than 200 kW [S6].

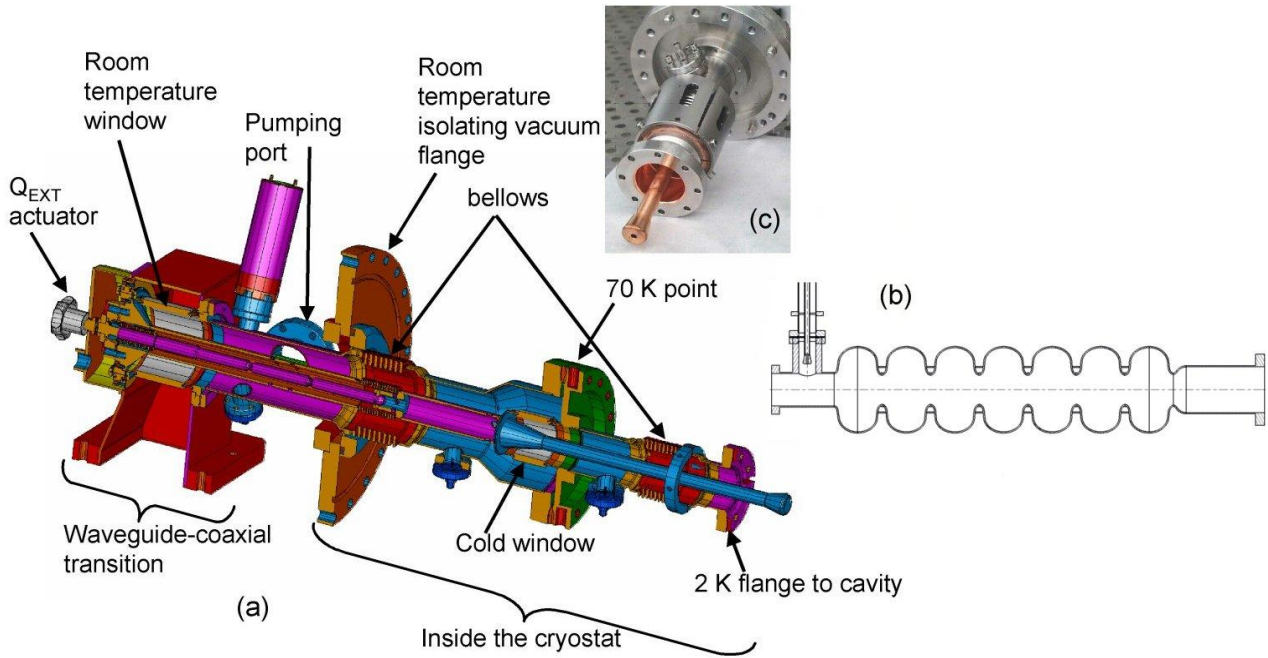

**Fig. 14:** Coaxial coupler of the TESLA cavity [S8]

The mechanical drawing allows us to illustrate many critical components in the design of the coaxial couplers for superconducting cavities. As clearly shown in the figure the coupler integrates several components:

*a) Waveguide to coaxial transition.*

*b) Bellows for $Q_{EXT}$ tunability.* For many accelerators it is necessary to tune the coupling by changing the penetration of the antenna in the pipe. In general $Q_{EXT}$ can vary from about $10^7$ to about $10^5$ depending on the application, and to change such coupling coefficient the coupler has to be inserted (or extracted) by several mm at operating frequencies around 1 GHz. This penetration has to be compensated by the bellows.

*c) Vacuum barriers (windows).* They prevent contamination of the SC structure and are made, in general, from $Al_2O_3$. These barriers are necessary also in normal-conducting accelerators but the requirement on the quality of the vacuum and reliability of the windows is much more stringent for SC structures. For this reason more than one vacuum barrier can be integrated in the coupler. In the example shown there are two windows. The first one is integrated in the waveguide-to-coaxial transition and is at room temperature, the second is a cold one. From the RF point of view the windows must be designed to be transparent to the electromagnetic field. Coaxial windows can be of several geometries: planar, cylindrical, or conical. Active pumping near the windows is desirable to avoid discharge problems during out-gassing events associated with varying power levels. As discussed in Section 6, multipacting phenomena can occur at the window location and can be particularly dangerous since a large amount of power can be deposited in small areas of the dielectric, leading to potential failures. Careful choice of geometry and coating with low-secondary-electron-emission-coefficient materials can mitigate this phenomenon. Another potential problem for RF windows is the exposure to radiation that can also lead to charging phenomena at the window surface and to damage of the window itself. Geometrical protection and metallic films of proper thickness can be used to decrease the incidence of this problem.

*d) Thermal barrier.* The RF power must be fed into the cold superconducting cavity and in the coupler we cross the boundary between the room-temperature and the low-temperature environment. This aspect imposes very tight requirements on geometries and a delicate balance between static and dynamic heat loads placed on the refrigerator system. Special thermal barriers have to be integrated in the coupler which must avoid the heating of the superconducting part of the cavity. Thermal simulations are also necessary for a complete and safe design of the coupler.

*e) Actuator for $Q_{EXT}$ tuning*. To tune $Q_{EXT}$ the penetration of the inner conductor is changed by an external actuator.

In the coupler, pumping ports for vacuum are also inserted as in the waveguides or coaxial for normal-conducting structures.

Waveguide coupling is conceptually simpler, since it does not require a transition between the waveguide, which usually carries the output power of the RF sources, and the cavity. Figure 15 shows two pictures of the CESR-B single-cell cavity with waveguide coupling [S9]. On the other hand, the size of the coupler is generally larger than the coaxial one and, because of this larger size, the contribution to the heating of the cryogenic environment is usually larger. Ceramic windows are also integrated in the waveguide but generally they are more difficult to manufacture than the coaxial ones.

As already discussed, the coupling strength can be changed by changing the size of the coupling iris or the longitudinal location of the waveguide with respect to the cavity end-cell, or by changing the location of the terminating short of the waveguide itself. Multipacting occurs in waveguides as well and can be moderated by the use of magnetic field biasing. A good review of some of these issues can be found in Ref. [S10].

In case of CW operation the requirements for higher average power are demanding for the design of a cooling system. Usually the central antenna and the bellows can be water, gas, or air cooled. Attention must also be paid to the thermal characteristics of the gaskets if the flange region proves to be a 'hot zone'. For certain materials (like aluminium, for example) it is possible to have vacuum leaks starting from ~150 degrees. In this case copper gaskets are recommended.

The last important parameter to be controlled (as in the case of NC cavities) is the transverse kick induced by the coupler and the multipole field components that can deteriorate the beam quality.

The remedy is to compensate this effect by alternating the coupler insertion on both sides of the beam propagation axis and, in the design phase, by reducing the ratio between the coupler and cavity diameters. Of course this problem is more serious for SC RF guns and for low-energy beams [BD3].

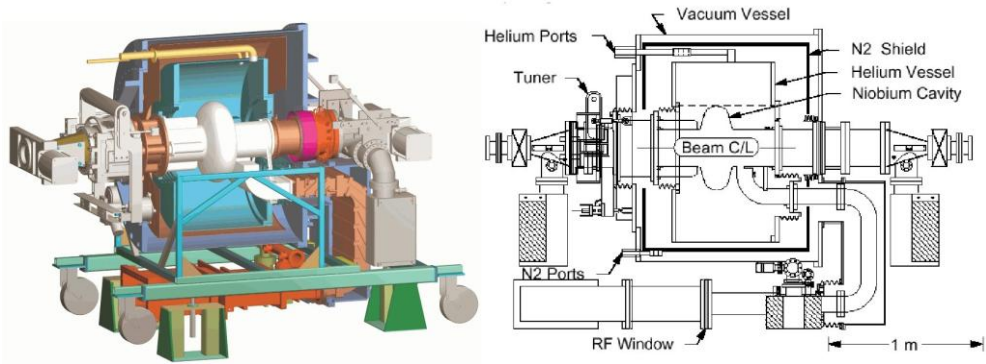

**Fig. 15:** CESR-B single-cell cavity with waveguide coupling [S9]

## 5    Couplers for travelling-wave structures

Travelling-wave structures are normal-conducting structures used for electron acceleration. They have an input coupler, many accelerating cells (~80), and an output coupler, as shown in Fig. 16. Because of the high gradient (from ~25 MV/m in S Band up to 80–100 MV/m in X Band) and high input power (~100 MW) the structures are fed by waveguides. Several types of coupler have been proposed for TW structures. They can be divided into two big families: the slot-type coupler, as shown in Fig.16, and the mode converter coupler [T1], shown in Fig.17.

In the first case the coupler is realized by connecting the waveguide and the first accelerating cell through a slot on the waveguide, as illustrated in Fig. 16. The travelling-wave accelerating mode ($TM_{01}$-like) is magnetically excited by the $TE_{10}$ mode of the waveguide. Matching is obtained by tuning the radius of the first accelerating cell ($R_c$) and the slot aperture ($w$). Electromagnetic codes are, in general, used to design these couplers and a possible tuning procedure is described in the final section of this paper.

In the second type of coupler the $TE_{10}$ mode of the coupler is converted into the $TM_{01}$ mode of the circular waveguide and this circular mode is then converted into the $TM_{01}$-like mode of the structure. The sketch of this coupler is given in Fig. 17 (a). The first matching is obtained by tuning the dimension of the two bumps, for example, while the circular waveguide mode is converted into the accelerating mode by tuning the radius and the iris aperture of the first accelerating cell. A more compact version of such a coupler is the waveguide coupler [shown in Fig. 17 (b)]. In this case the waveguide is directly connected to the accelerating section and the $TE_{10}$ mode is converted into the accelerating mode by tuning the first iris aperture and the radius of the first cell.

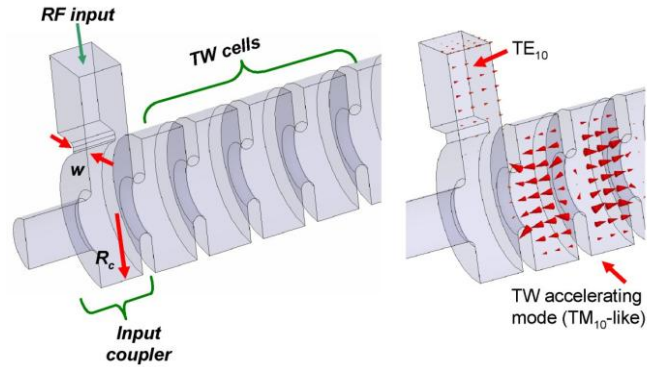

**Fig. 16:** Slot-type coupler for a TW structure

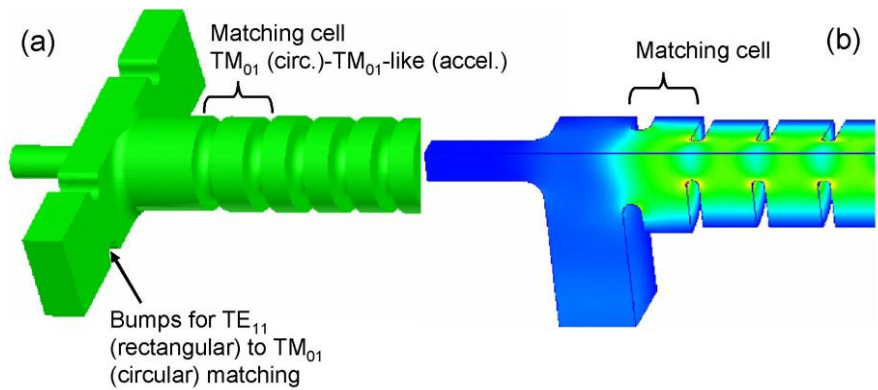

**Fig. 17:** (a) Mode converter coupler; (b) waveguide coupler [T1

The different couplers have advantages and drawbacks. The first one is of course that of being more compact since just one accelerating cell is sacrificed for the matching. On the other hand, a high magnetic field can occur in the slot region if it is not sufficiently rounded and, if only one slot is used for coupling, a strong dipole component of the field can occur in the coupler cell. Couplers of the second type are less compact, they need a splitter for symmetric feed but they completely cancel the problem of high magnetic field and pulsed heating.

Let us go into the details of such problems.

## 5.1 Field distortions

As observed in the case of SW cells, the couplers for TW sections introduce asymmetries in the field distribution in terms of multipole components of the field. Let us consider the case of slot-type couplers. If only one slot is used, the most important distortion is given by the dipole term. Several compensation techniques are possible: they are illustrated in Fig. 18.

The first one [Fig. 18(a)] is to introduce a transverse offset of the coupling cell in order to make the magnetic centre coincident with the beam axis [T2]. This technique does not give a perfect compensation of the kick in the sense that power still flows into the structure from one unbalanced side and the imaginary and real parts of the dipole field cannot be perfectly compensated at the same point. The second solution is the dual feed [Fig. 18(b)]. In this case the dipole component is perfectly compensated but, of course, one needs a splitter, which complicates the design. Possible solutions with a splitter integrated in the coupler have also been suggested [T3]. The last possibility of dipole kick compensation is the use of the so-called J-type coupler, proposed a few years ago [T4]. The sketch of this coupler is given in Fig. 18(c). The waveguide follows the profile of the coupler cell and two slots are opened on opposite sides. The dimensions and lengths of the waveguide around the coupler are tuned to have a perfect field in phase on the two slots.

In the case of mode converter couplers the dipole component is completely compensated but a stronger quadrupole component of the field can occur [T5]. In such a type of coupler, in fact, the rectangular geometry of the waveguide is directly inserted on the circular geometry of the accelerating field, as sketched in Fig. 19, and multipole components of the field can occur. A careful evaluation of such components must be done in order to investigate the effect on the beam dynamics. The integration of a splitting system in the waveguide coupler, as shown in Fig. 20, is easier.

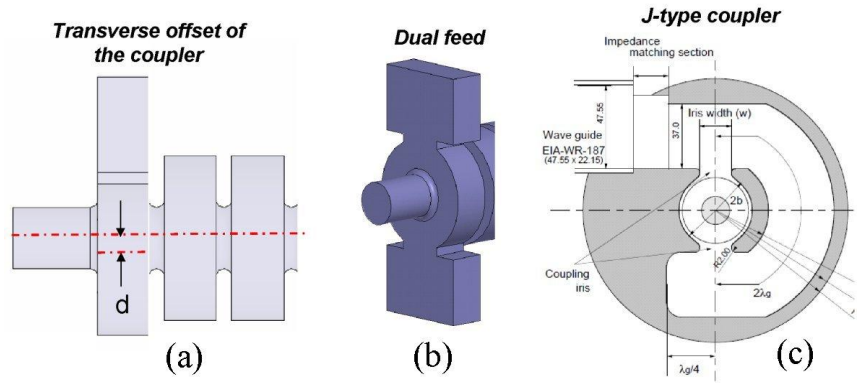

**Fig. 18:** Possible compensation techniques for dipole kick compensation in slot-type couplers for TW structures

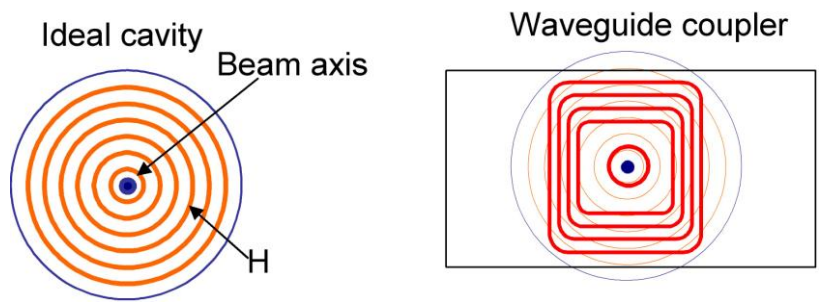

**Fig. 19:** Sketch of the magnetic field lines on a mode converter coupler

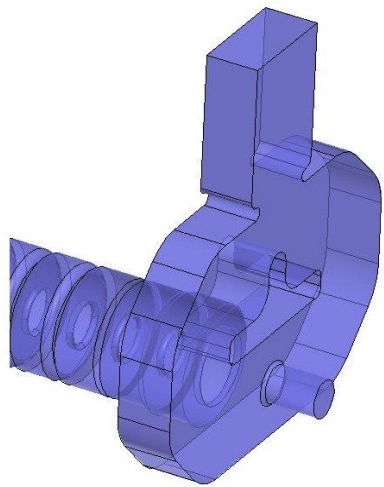

**Fig. 20:** Sketch of the integration of the power splitting system in a waveguide coupler

### 5.1 High magnetic field and pulse heating

Sharp edges in slot-type couplers can give a strong increase of the surface magnetic field. The criteria to reduce this magnetic field are similar to those discussed in Section 3 and they are based on a strong coupler rounding.

For the mode converter coupler this problem is completely overcome and the pulsed heating of the coupler is similar to that of the single cell of the structure.

## 6 Multipacting in couplers

Multipacting is a phenomenon of resonant electron multiplication. Several theoretical approaches have been developed to simulate this phenomenon [M1–M6].

Electrons are emitted from the walls because of the presence of high electric field. At a specific level of input power (field) the electrons can be accelerated, hit another wall (or the same wall) and force the emission of more electrons. If the Secondary Emission Yield (SEY) is bigger than 1, a large number of electrons can build up an electron avalanche, leading to remarkable power losses and heating of the walls, so that it becomes impossible to increase the cavity fields by increasing the incident power and damage to the surfaces and materials can also appear.

Multipacting is strongly enhanced in couplers by the presence of the ceramic windows because ceramic materials have a high SEY that stimulates the multipacting activity. Also bellows can drive multipacting because of the very high field zones. In coaxial couplers the multipactor threshold varies following a $(f_{RF}D)^4$ or a $ZD^4$ law where $f_{RF}$ is the frequency, $Z$ is the coaxial impedance, and $D$ the external diameter of the coaxial coupler.

Following these criteria and by a proper electromagnetic design it is possible to find shapes and configurations that minimize the multipacting activity.

To fight the multipacting phenomena, bias voltage on the central inner of the coaxial couplers can be implemented in order to shift the resonant condition [M8]. In waveguides the same effect is obtained by applying a magnetic field. Coating (of some tenths of nm) on ceramic RF windows with a low SEY material (usually Ti or TiN) can also be implemented and, in some cases, is mandatory [M9, M10].

## 7 Design techniques using electromagnetic codes

The design of couplers for SW and TW structures is performed, nowadays, using 3D electromagnetic codes. In the two cases the methods are different and, in the following, we discuss a couple of possible design techniques.

### 7.1 Design of couplers for travelling-wave structures

Several techniques have been proposed for TW structure design [D1, D3, TW1]. Let us consider the case of a slot-type coupler. In this case the dimensions of the slot and of the coupler cell have to be designed so as to minimize the reflected power at the waveguide input/output ports. Several geometrical parameters can be used for this purpose. The most effective are the radius of the cell and the width of the slot. Since by e.m. codes it is not possible to consider an infinite number of TW cells, in order to design the coupler one has to consider a TW structure with input and output couplers and a few TW cells. In this case it is possible to design the couplers by changing their dimensions minimizing the reflection coefficient at the waveguide input port and verifying that also the phase advance per cell in the TW structure is constant and equal to the nominal one. There are, in fact, cases in which the reflections at the input and output ports compensate each other and one has a minimum reflection coefficient even if the couplers are mismatched. This procedure is, in general, very time consuming.

Another technique is given in Ref. [D1] and is based on the equivalent circuit shown in Fig. 21. In the figure the two port networks with scattering matrix [*S*] correspond to the couplers that match the input/output waveguides to the disc loaded structure. Each cell of the TW structure is modelled by a two-port network. A perfectly matched coupler has the first element of the coupler scattering matrix ($S_{11}$) equal to zero. Based on this equivalent circuit and neglecting the losses in the coupler and cell it is possible to demonstrate that

$$S_{11} = 0 \Leftrightarrow \frac{\Gamma_s(n+2)}{\Gamma_s(n+1)} = \frac{\Gamma_s(n+1)}{\Gamma_s(n)} = e^{-j2\varphi} \quad \text{(with} \quad |\Gamma_s(n)| = 1) \quad , \tag{25}$$

where $\Gamma(n)$ is the reflection coefficient at the coupler input port when the structure is short circuited ($n$ is the position of the short-circuited cell) and $\phi$ is the phase advance per cell in the TW structure. A typical situation of short-circuited cells is shown in Fig. 22. Without going into details, widely discussed in Ref. [D1], in order to tune the coupler it is enough to vary only two of the input coupler dimensions (two parameters) until Eq. (25) is satisfied. This technique is much easier and faster since only one port and few cells need to be simulated.

As an example, the final reflection coefficient at the structure input port for a seven-cell X-Band structure is reported in Fig. 23. The coupler in this example has been tuned at 11.424 GHz. The finite number of reflection coefficient minima is given by the resonant SW patterns generated in the structure by the reflections at the input and output couplers. In fact, the coupler has a finite bandwidth (typically 50 MHz at 11 GH$_Z$) and, over this, there are reflections of the travelling wave from the input and output couplers. The minima are located in the pass-band of the periodic structure and their number is equal to the number of cells. By increasing the number of cells we progressively increase the number of minima in the pass-band.

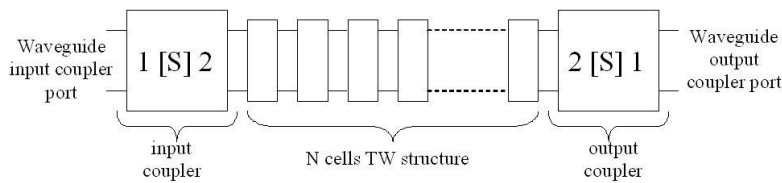

**Fig. 21:** Equivalent circuit of a TW structure with couplers

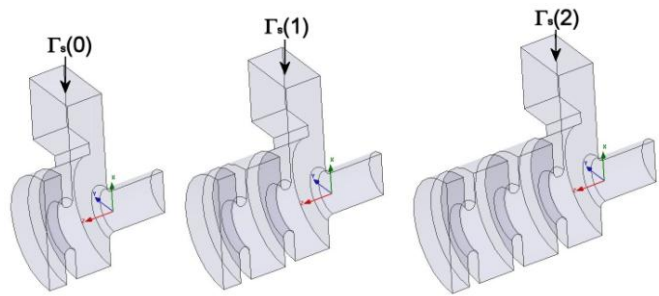

**Fig. 22:** Short-circuited cells for coupler matching

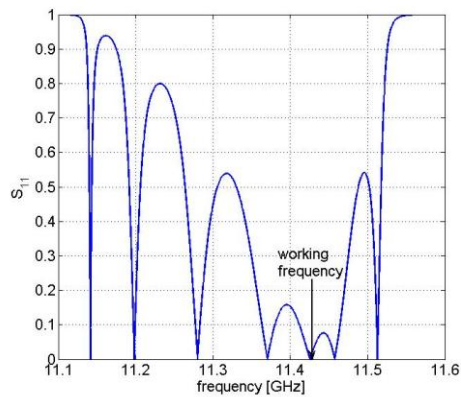

**Fig. 23:** Reflection coefficient at the coupler as a function of frequency in the case of a seven-cell structure operating at 11 GHz

## 7.2 Design of couplers for normal-conducting standing-wave structures

In this case one has to design the coupling slot in order to obtain the desired coupling coefficient without modifying the accelerating field distribution and the resonant frequency of the structure.

Let us consider the case of a multicell structure designed to operate at a given frequency with a certain field flatness [Fig. 24 (a)]. The insertion of the coupler in the central cell of the structure [Fig. 24(b)] detunes the coupling cell thus changing the resonant frequency and the field flatness. The waveguide input coupler, in fact, detunes the coupler cell because it increases its volume. To retune the structure one has to reduce the radius of the coupling cell itself [Fig. 24 (c)].

Since it is not possible to evaluate the coupling coefficient before the retuning of the coupling cell (because it depends on the field level into the coupling cell itself), in order to design the coupler one has to follow an iterative procedure:

1. fix the slot dimension,
2. simulate the structure retuning the coupler cell,
3. calculate the coupling coefficient,
4. if the coupling is not the desired one, start again from 1.

To simplify the design it is possible to simulate the coupler cell only with the proper boundary conditions (perfect $H$ for the accelerating $\pi$ mode), as given in Fig. 25. Looking at the definition of the coupling coefficient, one has to tune the slot and the radius of the cell in order to have a coupling coefficient equal to $N$ times ($N$ = total number of accelerating cells in the full structure) the desired coupling coefficient with a resonant frequency exactly equal to the resonant frequency of the structure without coupler.

As an example the reflection coefficient at the coupler input port for the seven-cell structure is given in Fig. 26 as a function of frequency. The different minima are the SW modes of the structure that can be excited from the coupler, and the couplings to these modes are, of course, different since they depend on the mode configuration. The working one has, in this example, $\beta = 1$.

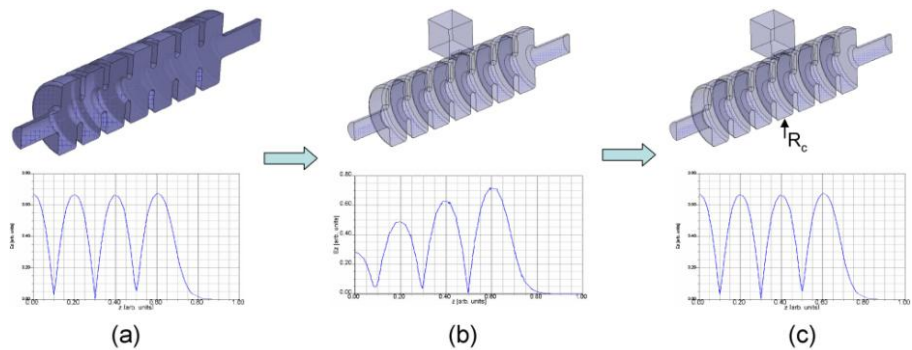

**Fig. 24:** Steps in the design of a coupler for a SW multicell cavity

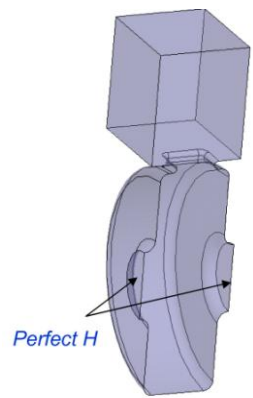

**Fig. 25:** Coupler cell of a multi-cell structure with the proper boundary conditions for accelerating $\pi$ mode

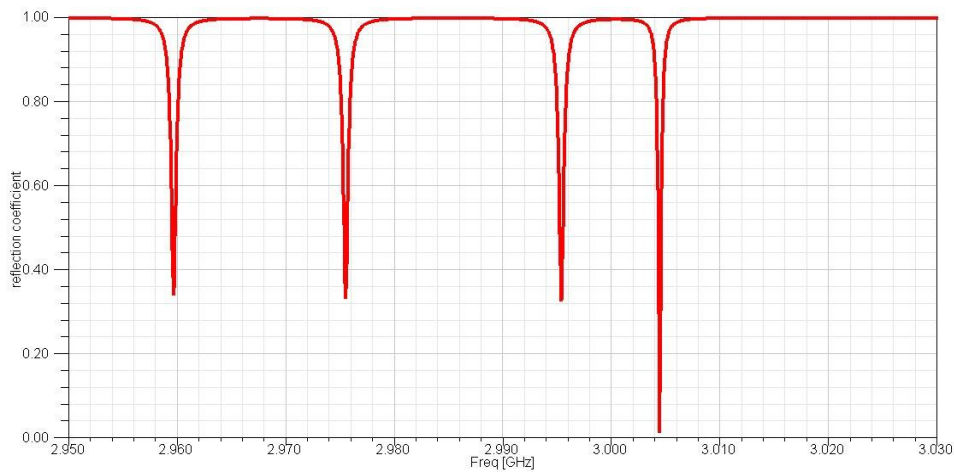

**Fig. 26:** Reflection coefficient at the coupler input port for the seven-cell SW structure

**Acknowledgements**

As mentioned in the abstract, the paper brings together pictures, data, and information from several works reported in the references and I would like to thank all the authors of the papers. I would like also to thank Miro Preger for careful revision of the paper.

**Subject bibliography**